\documentclass[12pt]{article}
\usepackage{amssymb}
\usepackage{amsmath}

\usepackage[normalem]{ulem}

\numberwithin{equation}{section}
\newcommand{\be}{\begin{eqnarray}}
\newcommand{\ee}{\end{eqnarray}}
\newcommand{\non}{\nonumber}
\newcommand{\id}{\mathbb{I}}
\newcommand{\tr}{\mathop{\rm tr}\nolimits}
\newcommand{\str}{\mathop{\rm str}\nolimits}
\newcommand{\diag}{\mathop{\rm diag}\nolimits}
\newcommand{\am}{\mathop{\rm am}\nolimits}
\newcommand{\cn}{\mathop{\rm cn}\nolimits}
\newcommand{\sn}{\mathop{\rm sn}\nolimits}
\newcommand{\dn}{\mathop{\rm dn}\nolimits}
\newcommand{\K}{{\rm K}}

\usepackage[usenames,dvipsnames]{color}

\begin{document}

\begin{titlepage}
\strut\hfill UMTG--284
\vspace{.5in}
\begin{center}

\LARGE Bethe ansatz for an AdS/CFT open spin chain\\
with non-diagonal boundaries\\
\vspace{1in}
\large Xin Zhang${}^{a}$,
Junpeng Cao${}^{a,b}$,
Shuai Cui${}^{a}$,
Rafael I. Nepomechie${}^{c}$, \\
Wen-Li Yang${}^{d,e}$,
Kangjie Shi${}^{d}$,
Yupeng Wang${}^{a,b}$\\
\vspace{1in}
\normalsize
{\it ${}^a$ Beijing National Laboratory for Condensed Matter
          Physics, Institute of Physics, Chinese Academy of Sciences, Beijing
           100190, China\\
${}^b$ Collaborative Innovation Center of Quantum Matter, Beijing, China\\
${}^c$ Physics Department, P.O. Box 248046, University of Miami,
Coral Gables, FL 33124, USA\\
${}^d$ Institute of Modern Physics, Northwest University, Xian 710069, China \\
${}^e$ Beijing Center for Mathematics and Information Interdisciplinary Sciences, Beijing, 100048,  China}
\end{center}

\vspace{.5in}

\begin{abstract}
    We consider the integrable open-chain transfer matrix
    corresponding to a $Y=0$ brane at one boundary, and a $Y_{\theta}=0$
    brane (rotated with the respect to the former by an angle
    $\theta$) at the other boundary.  We determine the exact eigenvalues of
    this transfer matrix in terms of solutions of a corresponding set
    of Bethe equations.
\end{abstract}

\end{titlepage}

\setcounter{footnote}{0}

\section{Introduction}

Two remarkable conjectures have commanded considerable attention for
over a decade: the AdS${}_{5}$/CFT${}_{4}$ correspondence \cite{Maldacena:1997re, Aharony:1999ti},
positing the equivalence of type IIB superstring theory on
$AdS_{5}\times S^{5}$ \cite{Metsaev:1998it} and ${\cal N}=4$
supersymmetric $SU(N)$ Yang-Mills theory in 3+1 dimensions
\cite{Brink:1976bc}; and the integrability of the spectral problem in
planar AdS${}_{5}$/CFT${}_{4}$ \cite{Beisert:2010jr}, positing that the energies of
string states, or equivalently, the scaling dimensions of all local
gauge-invariant operators in the planar limit of the dual gauge
theory, are described by an integrable 1+1 dimensional
model.\footnote{Integrability is believed to appear
also for AdS${}_{4}$/CFT${}_{3}$ \cite{Klose:2010ki} and AdS${}_{3}$/CFT${}_{2}$
\cite{Sfondrini:2014via}. However, we focus here on
AdS${}_{5}$/CFT${}_{4}$, which is the simplest and best-understood
case.}  We shall refer to the latter as the AdS/CFT integrable model.

This AdS/CFT integrable model is essentially the string world-sheet
quantum field theory in a light-cone gauge. It has a centrally-extended
$su(2|2)$ symmetry \footnote{Actually, the symmetry consists of two
copies of this algebra, but we focus here on just one copy.}, and the
spectrum includes four fundamental particles: two bosons, and two
fermions. The exact (non-relativistic) dispersion relation is known,
as are exact bulk and boundary world-sheet $S$-matrices.

The momentum quantization condition for a set of such particles on a
ring (i.e., periodic boundary conditions) of finite length leads
\cite{Beisert:2006qh, Martins:2007hb} to the all-loop asymptotic Bethe
equations \cite{Beisert:2005fw}, which determine the energies of
closed strings/scaling dimensions of single-trace operators in the
dual gauge theory, up to wrapping (finite-size) corrections.

Similarly, the momentum quantization condition for a set of such
particles on an {\it interval} of finite length leads to all-loop
asymptotic Bethe equations that determine the energies of {\it open}
strings/scaling dimensions of {\it determinant-like} operators in the
dual gauge theory, again up to wrapping (finite-size) corrections.
The detailed results depend on the specific boundary conditions at the
two ends of the interval.  Among the integrable cases that have been
studied are $Y=0$ branes \cite{Berenstein:2005vf, Hofman:2007xp} at both ends
\cite{Galleas:2009ye, Bajnok:2010ui, Bajnok:2012xc}; and $Y=0$ at one end and $\bar
Y=0$ at the other end \cite{Bajnok:2013wsa}. (For a review of
integrable boundary conditions in AdS/CFT, see \cite{Zoubos:2010kh}.)

The key technical step in deriving the various asymptotic Bethe
equations is to determine the eigenvalues of the corresponding
integrable inhomogeneous transfer matrices, which are constructed with
the bulk and -- for cases with boundaries -- boundary $S$-matrices.
The boundary $S$-matrices for $Y=0$ and $\bar Y=0$ branes are
diagonal.  However, the boundary $S$-matrix for a $Y_{\theta}=0$ brane
\cite{Bajnok:2013wsa}, which interpolates between them, is not
diagonal.  Hence, the problem of diagonalizing the transfer matrix
constructed with the latter boundary $S$-matrix is nontrivial, and is
the main goal of this paper.  Our strategy is to exploit the unbroken
$u(1)$ symmetry by carrying out the first step
of the nested algebraic Bethe ansatz, following
\cite{Martins:1997, Guan:1999}.  This leads to a
second-level open-chain spin-1/2 XXX transfer matrix with non-diagonal
boundary terms, which we diagonalize by introducing an inhomogeneous
term in its T-Q equation \cite{Cao:2013qxa, Wang2015}. A similar
strategy was employed to solve the open Hubbard \cite{Li:2013vdh} and
supersymmetric t-J \cite{Zhang:2013pha} models with
non-diagonal boundary interactions; however, those works used
coordinate Bethe ansatz
(instead of nested algebraic Bethe ansatz) for the first step.

The paper is organized as follows.  In section \ref{sec:prelim} we
introduce our notations, recall the relevant AdS/CFT bulk and boundary
$S$-matrices, and review the construction of the corresponding
integrable open-chain transfer matrix.  In section
\ref{sec:diagonalization} we determine the exact eigenvalues of this
transfer matrix in terms of solutions of a corresponding set of Bethe
equations.  We then use the unbroken $su(2)$ symmetry to derive
formulas for the number of distinct eigenvalues (and hence, number of
solutions of the Bethe equations) and their degeneracies.  We check
these results numerically for small system size.  In section
\ref{sec:discussion} we briefly discuss our results and note some
remaining problems.  In the appendix we propose a
generating functional for the eigenvalues of transfer matrices whose
auxiliary spaces belong to higher-dimensional representations of
$su(2|2)$.

\section{Construction of the transfer matrix}\label{sec:prelim}

Here we introduce our notations, recall the relevant AdS/CFT bulk and
boundary $S$-matrices, and review the construction of the
corresponding integrable open-chain transfer matrix.

\subsection{Parametrization}\label{sec:parametrization}

Following Arutyunov and Frolov \cite{Arutyunov:2007tc}, we use the elliptic
parametrization for the momentum $p$ and the
parameters $x^{\pm}$ \footnote{We generally
follow the conventions in {\it Mathematica} for the
Jacobi elliptic functions, e.g. $\cn (z,k) = {\rm JacobiCN}[z,k]$ and
$\K(k)={\rm EllipticK}[k]$. The one exception is $\am (z,k) = -i
{\rm Log}[{\rm JacobiCN}[z,k]+i {\rm JacobiSN}[z,k]]$, which is consistent with both
(\ref{xpxm}) and (\ref{pcrossing}).}
\be
p(z) = 2 \am(z,k) \,, \qquad
x^{\pm}(z)=\frac{1}{2g}\left(\frac{\cn (z,k)}{\sn  (z,k)}\pm
i\right)\left(1+\dn(z,k)\right)\,, \qquad k = - 4g^{2} \,,
\ee
such that
\be
\frac{x^{+}}{x^{-}}=e^{i p} \,,
\label{xpxm}
\ee
and
\be
x^{+} + \frac{1}{x^{+}}-x^{-} - \frac{1}{x^{-}} = \frac{2i}{g} \,.
\ee
(We shall often refrain from exhibiting the dependence of
$x^{\pm}$ and $p$ on the uniformizing parameter $z$.)
The parameter $g$ is the coupling constant of the AdS/CFT integrable
model (string tension), which is related to the 't Hooft
coupling $\lambda$ of the dual gauge theory by
$g=\sqrt{\lambda}/(2\pi) >0$.

The two periods are given by
\be
2\omega_{1} = 4 \K(k)\,, \qquad
2\omega_{2} = 4i \K(1-k) - 4 \K(k)\,,
\ee
where $\K(k)$ is the complete elliptic integral of the first
kind. The crossing transformation is effectuated with a shift of $z$
by the half-period $\omega_{2}$,
\be
x^{\pm}(z+\omega_{2}) &=& \frac{1}{x^{\pm}(z)} \,, \\
p(z+\omega_{2}) &=& -p(z) \label{pcrossing} \,, \\
E(z+\omega_{2}) &=& -E(z)  \,,
\ee
where
$E(z)=-\frac{ig}{2}\left(x^{+}-\frac{1}{x^{+}}-x^{-}+\frac{1}{x^{-}}\right) =
\dn(z,k)$ is the energy. We note that $z \mapsto -z$ corresponds to a
reflection,
\be
x^{\pm}(-z) = - x^{\mp}(z)\,, \qquad p(-z) = -p(z) \,, \qquad E(-z) =
E(z) \,.
\ee
We define $u(z)$ by
\be
x^{\pm} + \frac{1}{x^{\pm}} = \frac{2}{g}\left(u \pm
\frac{i}{2}\right)\,,
\label{udef}
\ee
and therefore
\be
u(z+\omega_{2})=u(z)\,, \qquad u(-z)=-u(z) \,.
\ee

\subsection{$S$-matrices}\label{sec:Smatrix}

As already noted, there are four fundamental particles.
Let us denote the corresponding Zamolodchikov-Faddeev operators by
$A_{i}^{\dagger}(z)\,, i = 1, 2, 3, 4$, where $i=1, 2$
are bosonic and $i=3, 4$ are fermionic. The 
matrix elements of the bulk $S$-matrix are defined by
\be
A_{i}^{\dagger}(z_{1})\, A_{j}^{\dagger}(z_{2}) = 
\sum_{i', j' = 1}^{4}
S^{\, i' j'}_{i\, j}(z_{1},z_{2})\, A_{j'}^{\dagger}(z_{2})\, 
A_{i'}^{\dagger}(z_{1})\,,
\ee
which can be arranged into a $16 \times 16$ matrix as follows
\be
S(z_1,z_2) = \sum_{i, i', j, j' = 1}^{4} S^{\, i' j'}_{i\, j}(z_1,z_2)\, 
e_{i\, i'}\otimes e_{j\, j'}\,, \qquad \left( e_{ij} \right)_{ab} =
\delta_{a,i} \delta_{b,j} \,.
\ee
We work with a graded version of Beisert's $su(2|2)$
$S$-matrix \cite{Beisert:2005tm}. Specifically, following
Arutyunov and Frolov \cite{Arutyunov:2008zt}, we take
\be
S(z_1,z_2) = \sum_{k=1}^{10}a_{k}(z_1,z_2) \Lambda_{k}\,,
\label{bulkS}
\ee
where the matrices $\Lambda_{1}\,, \ldots\,, \Lambda_{10}$
are given in terms of quantities $E_{kilj}$ defined by
\be
E_{kilj} = e_{ki} \otimes e_{lj} \,.
\label{Ekilj}
\ee

Hence, $S(z_1,z_2)$ has the following matrix structure
\be
\left(
\begin{array}{cccccccccccccccc}
 a_1 & 0 & 0 & 0 & 0 & 0 & 0 & 0 & 0 & 0 & 0 & 0 & 0 & 0 & 0 & 0 \\
 0 & \frac{a_1}{2}+\frac{a_2}{2} & 0 & 0 & \frac{a_1}{2}-\frac{a_2}{2} & 0 & 0 & 0 & 0 & 0 & 0 & a_7 & 0 & 0 &
   -a_7 & 0 \\
 0 & 0 & a_5 & 0 & 0 & 0 & 0 & 0 & a_9 & 0 & 0 & 0 & 0 & 0 & 0 & 0 \\
 0 & 0 & 0 & a_5 & 0 & 0 & 0 & 0 & 0 & 0 & 0 & 0 & a_9 & 0 & 0 & 0 \\
 0 & \frac{a_1}{2}-\frac{a_2}{2} & 0 & 0 & \frac{a_1}{2}+\frac{a_2}{2} & 0 & 0 & 0 & 0 & 0 & 0 & -a_7 & 0 & 0 &
   a_7 & 0 \\
 0 & 0 & 0 & 0 & 0 & a_1 & 0 & 0 & 0 & 0 & 0 & 0 & 0 & 0 & 0 & 0 \\
 0 & 0 & 0 & 0 & 0 & 0 & a_5 & 0 & 0 & a_9 & 0 & 0 & 0 & 0 & 0 & 0 \\
 0 & 0 & 0 & 0 & 0 & 0 & 0 & a_5 & 0 & 0 & 0 & 0 & 0 & a_9 & 0 & 0 \\
 0 & 0 & a_{10} & 0 & 0 & 0 & 0 & 0 & a_6 & 0 & 0 & 0 & 0 & 0 & 0 & 0 \\
 0 & 0 & 0 & 0 & 0 & 0 & a_{10} & 0 & 0 & a_6 & 0 & 0 & 0 & 0 & 0 & 0 \\
 0 & 0 & 0 & 0 & 0 & 0 & 0 & 0 & 0 & 0 & a_3 & 0 & 0 & 0 & 0 & 0 \\
 0 & a_8 & 0 & 0 & -a_8 & 0 & 0 & 0 & 0 & 0 & 0 & \frac{a_3}{2}+\frac{a_4}{2} & 0 & 0 &
   \frac{a_3}{2}-\frac{a_4}{2} & 0 \\
 0 & 0 & 0 & a_{10} & 0 & 0 & 0 & 0 & 0 & 0 & 0 & 0 & a_6 & 0 & 0 & 0 \\
 0 & 0 & 0 & 0 & 0 & 0 & 0 & a_{10} & 0 & 0 & 0 & 0 & 0 & a_6 & 0 & 0 \\
 0 & -a_8 & 0 & 0 & a_8 & 0 & 0 & 0 & 0 & 0 & 0 & \frac{a_3}{2}-\frac{a_4}{2} & 0 & 0 &
   \frac{a_3}{2}+\frac{a_4}{2} & 0 \\
 0 & 0 & 0 & 0 & 0 & 0 & 0 & 0 & 0 & 0 & 0 & 0 & 0 & 0 & 0 & a_3 \\
\end{array}
\right) \non 
\ee
and the matrix elements $a_{k} = a_{k}(z_1,z_2)$ are given by \cite{Arutyunov:2008zt}
\be
a_1 &=& 1\,, \non \\
a_2 &=& 2\,\frac{(x^+_1-x^+_2) (x^-_1
   x^+_2-1)x^-_2}{(x^+_1 - x^-_2)(x^-_1 x^-_2-1)
    x^+_2}-1\,, \non \\
a_3 &=& \frac{x^+_2-x^-_1}{x^-_2-x^+_1}
\frac{\tilde{\eta}_1
  \tilde{\eta}_2}{ \eta_1\eta_2} \,, \non \\
a_4 &=& \frac{(x^-_1-x^+_2)}{(x^-_2-x^+_1)}\frac{\tilde{\eta}_1
 \tilde{\eta}_2}{ \eta_1\eta_2} - 2\,\frac{
   (x^-_2 x^+_1-1) (x^+_1-x^+_2)
  x^-_1 }{(x^-_1 x^-_2-1)
   (x^-_2-x^+_1) x^+_1} \frac{\tilde{\eta}_1
  \tilde{\eta}_2}{ \eta_1\eta_2} \,, \non \\
a_5 &=& \frac{x^-_2-x^-_1}{x^-_2-x^+_1 } \frac{\tilde{\eta}_2}{\eta 
_2}  \,, \non \\
a_6 &=& 
\frac{x^+_1-x^+_2}{x^+_1-x^-_2}\frac{\tilde{\eta}_1}{\eta_1}\,, \non 
\\
a_7 &=& -\frac{i (x^-_1-x^+_1)
   (x^-_2-x^+_2)
   (x^+_1-x^+_2)}{(x^-_1
   x^-_2-1) (x^-_2-x^+_1) }\frac{1}{\eta
   _1 \eta _2}\,, \non \\
a_8 &=& \frac{i x^-_1 x^-_2
   (x^+_1-x^+_2)}{(x^-_1 x^-_2-1)
   (x^-_2-x^+_1) x^+_1
   x^+_2}\tilde{\eta}_1\tilde{\eta}_2 \,, \non \\
a_9 &=& \frac{x^+_1-x^-_1}{x^+_1-x^-_2}\frac{\tilde{\eta}_2}{\eta_1} 
\,, \non \\
a_{10} &=& \frac{x^-_2-x^+_2}{x^-_2-x^+_1}\frac{\tilde{\eta}_1}{\eta_2}\,,
\label{ak}
\ee
where $x^{\pm}_{i} = x^{\pm}(z_{i})$. Moreover,
\be
\eta_1 = e^{ip_2/2}\eta(z_1)\,,\quad 
\eta_2 =\eta(z_2)\,,\quad
\tilde{\eta}_1 = \eta(z_1)\,,\quad 
\tilde{\eta}_2 = e^{ip_1/2}\eta(z_2)\,,
\ee
where $p_{i} = p(z_{i})$ and 
\be
\eta(z)= \sqrt{\frac{2}{g}}\frac{\dn\, \frac{z}{2}\big(
\cn\, \frac{z}{2}+i \, \sn\, \frac{z}{2}\dn\,
\frac{z}{2}\big)}{1+4g^2\, {\sn^4\frac{z}{2}}}\,.
\ee

This $S$-matrix satisfies the graded Yang-Baxter equation
\footnote{We work in the so-called string (rather than
spin-chain) frame/basis, where the $S$-matrix obeys a standard (rather
than twisted) Yang-Baxter equation.}
\be
S_{12}(z_1,z_2)\, S_{13}(z_1,z_3)\, S_{23}(z_2,z_3) =
S_{23}(z_2,z_3)\, S_{13}(z_1,z_3)\, S_{12}(z_1,z_2) \,,
\label{YBE}
\ee
where $S_{12}(z_1,z_2) = S(z_1,z_2) \otimes \id\,, S_{13}(z_1,z_3) = {\cal
P}_{23} S_{12}(z_1,z_3) {\cal P}_{23}\,, S_{23}(z_2,z_3) = {\cal
P}_{12} S_{13}(z_2,z_3) {\cal P}_{12}$, and ${\cal P}$ denotes the
{\em graded} permutation matrix
\be
{\cal P} = \sum_{a,b=1}^{4}  (-1)^{\epsilon_{a} \epsilon_{b}} e_{ab} \otimes e_{ba} \,,
\ee
where the gradings are given by $\epsilon_{1}=\epsilon_{2}=0\,, \epsilon_{3}=\epsilon_{4}=1$.
As is well known, the $S$-matrix has $su(2)\oplus su(2)$ symmetry,
\be
\left[ S_{12}(z_{1}, z_{2})\,, G_{1}\, G_{2} \right] = 0 \,, \qquad
G = \left(\begin{array}{cc}
g_{L} & 0 \\
0 & g_{R}
\end{array}\right) \,,
\label{SU2SU2symmetry}
\ee
where $g_{L}$ and $g_{R}$ are independent $2 \times 2$ special
unitary matrices.

For the right boundary, we consider a boundary $S$-matrix that is
diagonal \cite{Hofman:2007xp, Ahn:2010xa},
\be
R^{R}(z) = \diag(e^{-i p/2}\,, -e^{i p/2} \,, 1\,, 1 ) \,,
\label{RBSM}
\ee
corresponding to a $Y=0$ brane \cite{Berenstein:2005vf, Hofman:2007xp}.
It satisfies the right boundary Yang-Baxter equation
\cite{Cherednik:1985vs, Sklyanin:1988yz, Ghoshal:1993tm}
\be
S_{12}(z_{1},z_{2})\, R^{R}_{1}(z_{1})\, S_{21}(z_{2},-z_{1})\,
R^{R}_{2}(z_{2}) = R^{R}_{2}(z_{2})\, S_{12}(z_{1},-z_{2})\,
R^{R}_{1}(z_{1})\, S_{21}(-z_{2},-z_{1})\,,
\label{RBYBE}
\ee
where $S_{21}(z_{1},z_{2})= {\cal P}_{12}\, S_{12}(z_{1},z_{2})\,
{\cal P}_{12}$, $R^{R}_{1}(z) = R^{R}(z) \otimes \id$
and $R^{R}_{2}(z) = {\cal P}_{12}\, R^{R}_{1}(z)\,
{\cal P}_{12}$.

For the left boundary, we consider a non-diagonal boundary $S$-matrix
\cite{Bajnok:2013wsa},
\be
R^{L}(z) = O^{t}(\theta) R^{R}(-z) O(\theta) \,,
\label{LBSM}
\ee
where $O(\theta)$ is the rotation matrix
\be
O(\theta) = \left(
\begin{array}{cccc}
\cos \theta &\sin \theta & 0 & 0 \\
-\sin \theta & \cos \theta & 0 & 0 \\
0 & 0 & 1 & 0\\
0 & 0 & 0 & 1
\end{array}\right) \,,
\ee
and $\theta$ is an arbitrary angle.
This boundary $S$-matrix, which corresponds to a $Y_{\theta}=0$ brane,
interpolates between $Y=0$ ($\theta=0$) and $\bar Y=0$
($\theta=\pi/2$).
This boundary $S$-matrix evidently preserves the right $su(2)$ symmetry
\be
\left[ R^{L}(z)\,, g_{R} \right] = 0 \,,
\ee
but breaks the left $su(2)$ symmetry.

\subsection{Transfer matrix}\label{sec:transfer}

The open-chain transfer matrix for a single copy of $su(2|2)$ is given
by \cite{Sklyanin:1988yz, Bracken:1997, Murgan:2008fs} \footnote{In
order to derive the AdS/CFT all-loop asymptotic Bethe equations, one
must also take into account the second copy of the $su(2|2)$
$S$-matrix.  However, the most difficult technical part of the derivation is the
diagonalization of the transfer matrix for a single copy, on which we
focus here.}
\be
t(z\,; \{z_{i}\}) = \str_{a} \left\{ R_{a}^{L}(z)\, T_{a}(z\,; \{z_{i}\}) \,
R_{a}^{R}(z)\,  \widehat T_{a}(z\,; \{z_{i}\}) \right\} \,,
\label{transfer}
\ee
where the monodromy matrices are given by
\be
T_{a}(z\,; \{z_{i}\}) &=& S_{a N}(z,z_{N}) \cdots S_{a 1}(z,z_{1}) \,,
\non \\
\widehat T_{a}(z\,; \{z_{i}\}) &=& S_{1 a}(z_{1}, -z) \cdots S_{N
a}(z_{N}, -z)\,,
\label{monodromy}
\ee
the auxiliary space is denoted by $a$,
and str denotes super trace. The $\{z_{i}\}$, which correspond to the
rapidities of the $N$ particles on an interval, are to be regarded as
fixed inhomogeneities. (To lighten the notation, we shall often
suppress the dependence on these inhomogeneities.)
By construction (see e.g. \cite{Sklyanin:1988yz, Bracken:1997}),
the transfer matrix has the fundamental
commutativity property
\be
\left[ t(z\,; \{z_{i}\}) \,, t(z'\,; \{z_{i}\}) \right] = 0
\label{commutativity}
\ee
for arbitrary values of $z$ and $z'$.  For the boundary $S$-matrices
(\ref{RBSM}) and (\ref{LBSM}) that we consider here, the transfer
matrix also has the right $su(2)$ symmetry
\be
\left[ t(z\,; \{z_{i}\}) \,, \vec S \right]  =0 \,,
\label{su2}
\ee
where
\be
\vec S  = \sum_{n=1}^{N} \vec S_{n} \,, \qquad
\vec S_{n} = \left(\begin{array}{cc}
0 & 0 \\
0 & \frac{1}{2}\vec \sigma
\end{array}\right)_{n} \,.
\ee

\section{Exact diagonalization of the transfer matrix}\label{sec:diagonalization}

We turn now to the main task of deriving the eigenvalues of the transfer matrix
(\ref{transfer}) and obtaining the corresponding Bethe equations.

\subsection{Nested algebraic Bethe ansatz}\label{sec:NABA}

The transfer matrix has an unbroken $u(1) \subset su(2)$
symmetry. In particular, the state with ``all spins down''
\be
|0\rangle=\otimes_{j=1}^N|0\rangle_j\,, \qquad
|0\rangle_j=\left(
              \begin{array}{c}
                0 \\
                0 \\
                0 \\
                1 \\
              \end{array}
            \right)_j
\label{reference}
\ee
is an eigenstate of the transfer matrix. Therefore, using this
state as the reference state, we can carry out the first step
of the nested algebraic Bethe ansatz, following
\cite{Martins:1997, Guan:1999}. To this end,
it is convenient to write the boundary $S$-matrices (\ref{RBSM}),
(\ref{LBSM}) as
\be
R^{R}(z)&=&\left(
           \begin{array}{cccc}
             K^-_1(z)&  &  &  \\
              & K^-_2(z) &  &  \\
              &  & 1 &  \\
              &  &  & 1 \\
           \end{array}
         \right)\,, \quad
R^{L}(z)=\left(
           \begin{array}{cccc}
             K^+_1(z)& K^+_2(z) &  &  \\
             K^+_3(z) & K^+_4(z) &  &  \\
              &  & 1 &  \\
              &  &  & 1 \\
           \end{array}
         \right)\,, \label{boundSmat}
\ee
where
\be
K^-_1(z)&=&e^{-i p(z)/2}\,, \quad  K^-_2(z) = -e^{i p(z)/2}\,, \non \\
K^+_1(z)&=&\cos^2\theta e^{ip(z)/2}-\sin^2\theta e^{-ip(z)/2}\,, \quad
K^+_2(z)=\sin\theta\cos\theta (e^{ip(z)/2}+e^{-ip(z)/2})\,, \non\\
K^+_3(z)&=&\sin\theta\cos\theta (e^{ip(z)/2}+e^{-ip(z)/2})\,, \quad
K^+_4(z)=\sin^2\theta e^{ip(z)/2}-\cos^2\theta e^{-ip(z)/2} \,.
\label{boundSmatelem}
\ee 	
We also write the monodromy matrices (\ref{monodromy}) as
follows
\be
T_{a}(z\,;\{z_i\})&=&\left(
         \begin{array}{cccc}
           A_{11}(z) & A_{12}(z) & E_1(z) & C_1(z) \\
           A_{21}(z) & A_{22}(z) & E_{2}(z) & C_{2}(z) \\
           C_{4}(z) & C_{5}(z) & D(z) & C_{3}(z) \\
           B_{1}(z) & B_{2}(z) & F(z) & B(z) \\
         \end{array}
       \right)\,, \\
\widehat{T}_{a}(z\,;\{z_i\})&=&\left(
         \begin{array}{cccc}
           \bar{A}_{11}(z) & \bar{A}_{12}(z) & \bar{E}_1(z) & \bar{C}_1(z) \\
           \bar{A}_{21}(z) & \bar{A}_{22}(z) & \bar{E}_{2}(z) & \bar{C}_{2}(z) \\
           \bar{C}_{4}(z) & \bar{C}_{5}(z) & \bar{D}(z) & \bar{C}_{3}(z) \\
           \bar{B}_{1}(z) & \bar{B}_{2}(z) & \bar{F}(z) & \bar{B}(z) \\
         \end{array}
       \right) \,.
\ee

\subsubsection{The action of the transfer matrix on the reference state}

We observe that the elements of the monodromy 
matrices have the following action on the reference state
\be
&&A_{11}(z)|0\rangle=A_{22}(z)|0\rangle=\prod_{i=1}^Na_5(z,z_i)|0\rangle\,,\\
&&D(z)|0\rangle=\prod_{i=1}^Na_{14}(z,z_i)|0\rangle,\quad B(z)|0\rangle=\prod_{i=1}^Na_{3}(z,z_i)|0\rangle\,,\\
&&A_{12}(z)|0\rangle=A_{21}(z)|0\rangle=0,\quad C_{j}(z)|0\rangle=0\,,\\
&&F(z)|0\rangle\neq0,\quad B_j(z)|0\rangle\neq0\,,\\
&&\bar{A}_{11}(z)|0\rangle=\bar{A}_{22}(z)|0\rangle=\prod_{i=1}^Na_6(z_i,-z)|0\rangle,\\
&&\bar{D}(z)|0\rangle=\prod_{i=1}^Na_{14}(z_i,-z)|0\rangle,\quad \bar{B}(z)|0\rangle=\prod_{i=1}^Na_{3}(z_i,-z)|0\rangle\,,\\
&&\bar{A}_{12}(z)|0\rangle=\bar{A}_{21}(z)|0\rangle=0,\quad \bar{C}_{j}(z)|0\rangle=0\,,\\
&&\bar{F}(z)|0\rangle\neq0,\quad \bar{B}_j(z)|0\rangle\neq0\,.
\ee
Here and below we use the following notations
\be
    a_{11}(z_1,z_2) = \tfrac{1}{2}(a_{1}(z_1,z_2) -
    a_{2}(z_1,z_2))\,, \qquad
    a_{12}(z_1,z_2) = \tfrac{1}{2}(a_{1}(z_1,z_2) +
    a_{2}(z_1,z_2))\,, \\
    a_{13}(z_1,z_2) = \tfrac{1}{2}(a_{3}(z_1,z_2) -
    a_{4}(z_1,z_2))\,, \qquad
    a_{14}(z_1,z_2) = \tfrac{1}{2}(a_{3}(z_1,z_2) +
    a_{4}(z_1,z_2))\,,
\ee
where $a_{1}\,, \ldots\,, a_{10}$ are given by (\ref{ak}).

The double-row monodromy matrix is defined as
\be
U_{a}(z)=T_{a}(z\,;\{z_i\})\, R^{R}_{a}(z)\, \widehat{T}_{a}(z\,;\{z_i\})=\left(
         \begin{array}{cccc}
           \mathcal{A}_{11}(z) & \mathcal{A}_{12}(z) & \mathcal{E}_1(z) & \mathcal{C}_1(z) \\
           \mathcal{A}_{21}(z) & \mathcal{A}_{22}(z) & \mathcal{E}_{2}(z) & \mathcal{C}_{2}(z) \\
           \mathcal{C}_{4}(z) & \mathcal{C}_{5}(z) & \mathcal{D}(z) & \mathcal{C}_{3}(z) \\
           \mathcal{B}_{1}(z) & \mathcal{B}_{2}(z) & \mathcal{F}(z) & \mathcal{B}(z) \\
         \end{array}
       \right),
\label{doublerow}
\ee
We obtain
\be
\mathcal{B}(z)|0\rangle&=&B(z)\bar{B}(z)|0\rangle,\\
\mathcal{A}_{11}(z)|0\rangle&=&K_1^-(z)A_{11}(z)\bar{A}_{11}(z)|0\rangle+C_1(z)\bar{B}_1(z)|0\rangle\,,\\
\mathcal{A}_{22}(z)|0\rangle&=&K_2^-(z)A_{22}(z)\bar{A}_{22}(z)|0\rangle+C_2(z)\bar{B}_2(z)|0\rangle\,,\\
\mathcal{A}_{12}(z)|0\rangle&=&C_1(z)\bar{B}_2(z)|0\rangle\,,\\
\mathcal{A}_{21}(z)|0\rangle&=&C_2(z)\bar{B}_1(z)|0\rangle\,,\\
\mathcal{D}(z)|0\rangle&=&K^-_1(z)C_4(z)\bar{E}_1(z)|0\rangle+K^-_2(z)C_5(z)\bar{E}_2(z)|0\rangle+C_3(z)\bar{F}(z)|0\rangle\non\\
&&+D(z)\bar{D}(z)|0\rangle \,.
\ee
In order to obtain the actions of operators $\mathcal{A}_{ij}(z)$ and
$\mathcal{D}(z)$ on the reference state,
we use exchange relations derived from the Yang-Baxter equation
(\ref{YBE})
\be
T_1(z\,;\{z_i\})\, S_{12}(z,-z)\, \widehat{T}_2(z\,;\{z_i\})=
\widehat{T}_2(z\,;\{z_i\})\, S_{12}(z,-z)\, T_1(z\,;\{z_i\})\,.
\label{relationTT}
\ee
After some algebra, we obtain 
\be
&&C_1(z)\bar{B}_1(z)|0\rangle=\frac{a_9(z,-z)}{a_3(z,-z)}\left[A_{11}(z)\bar{A}_{11}(z)-B(z)\bar{B}(z)\right]|0\rangle\,,\\
&&C_2(z)\bar{B}_2(z)|0\rangle=\frac{a_9(z,-z)}{a_3(z,-z)}\left[A_{22}(z)\bar{A}_{22}(z)-B(z)\bar{B}(z)\right]|0\rangle\,,\\
&&\bar{C}_1(z)B_1(z)|0\rangle=\frac{a_{10}(z,-z)}{a_3(z,-z)}\left[A_{11}(z)\bar{A}_{11}(z)-B(z)\bar{B}(z)\right]|0\rangle\,,\\
&&\bar{C}_2(z)B_2(z)|0\rangle=\frac{a_{10}(z,-z)}{a_3(z,-z)}\left[A_{22}(z)\bar{A}_{22}(z)-B(z)\bar{B}(z)\right]|0\rangle\,,\\
&&\mathcal{A}_{12}(z)|0\rangle=C_1(z)\bar{B}_2(z)|0\rangle=0\,,\\
&&\mathcal{A}_{21}(z)|0\rangle=C_2(z)\bar{B}_1(z)|0\rangle=0\,,\\
&&\mathcal{A}_{11}(z)|0\rangle=\left[K_1^-(z)+\frac{a_9(z,-z)}{a_3(z,-z)}\right]A_{11}(z)\bar{A}_{11}(z)|0\rangle-\frac{a_9(z,-z)}{a_3(z,-z)}B(z)\bar{B}(z)|0\rangle\,,\\
&&\mathcal{A}_{22}(z)|0\rangle=\left[K_2^-(z)+\frac{a_9(z,-z)}{a_3(z,-z)}\right]A_{22}(z)\bar{A}_{22}(z)|0\rangle-\frac{a_9(z,-z)}{a_3(z,-z)}B(z)\bar{B}(z)|0\rangle\,.
\ee

We define
\be
\tilde{\mathcal{A}}_{ij}(z)=\mathcal{A}_{ij}(z)+\delta_{ij}\frac{a_9(z,-z)}{a_3(z,-z)}\mathcal{B}(z)\,.
\ee
Then
\be
\tilde{\mathcal{A}}_{12}(z)|0\rangle&=&\tilde{\mathcal{A}}_{21}(z)|0\rangle=0\,,\\
\tilde{\mathcal{A}}_{11}(z)|0\rangle&=&\left[K_1^-(z)+\frac{a_9(z,-z)}{a_3(z,-z)}\right]A_{11}(z)\bar{A}_{11}(z)|0\rangle\non\\
&=&\frac{e^{ip(z)/2}+e^{-ip(z)/2}}{2}\prod_{k=1}^Na_5(z,z_k)a_6(z_k,-z)|0\rangle\,,\\
\tilde{\mathcal{A}}_{22}(z)|0\rangle&=&\left[K_2^-(z)+\frac{a_9(z,-z)}{a_3(z,-z)}\right]A_{22}(z)\bar{A}_{22}(z)|0\rangle\non\\
&=&-\frac{e^{ip(z)/2}+e^{-ip(z)/2}}{2}\prod_{k=1}^Na_5(z,z_k)a_6(z_k,-z)|0\rangle\,.
\ee
From the Yang-Baxter relation (\ref{relationTT}), we also obtain
\be
&&\left[C_4(z)\bar{E}_1(z)+a_{11}(z,-z)
C_5(z)\bar{E}_2(z)+a_{10}(z,-z) C_3(z) \bar{F}(z)\right]|0\rangle\non\\
&&=\left[\frac{a_{13}(z,-z)a_{10}(z,-z)}{a_3(z,-z)}
{B}(z)\bar{B}(z)+\frac{a_{10}(z,-z)a_{14}(z,-z)}{a_3(z,-z)}{A}_{11}(z) \bar{A}_{11}(z)-a_{10}(z,-z) D(z) \bar{D}(z)\right]|0\rangle\,,\non\\
&&\left[a_{11}(z,-z) C_4(z) \bar{E}_1(z)+C_5(z)
\bar{E}_2(z)+a_{10}(z,-z) C_3(z) \bar{F}(z)\right]|0\rangle\non\\
&&=\left[\frac{a_{13}(z,-z)a_{10}(z,-z)}{a_3(z,-z)}{B}(z)
\bar{B}(z)+\frac{a_{10}(z,-z)a_{14}(z,-z)}{a_3(z,-z)}{A}_{22}(z)
\bar{A}_{22}(z)-a_{10}(z,-z) D(z) \bar{D}(z)\right]|0\rangle\,,\non\\
&&\left[a_9(z,-z) C_4(z) \bar{E}_1(z)+a_9(z,-z) C_5(z)
\bar{E}_2(z)-C_3(z) \bar{F}(z)\right]|0\rangle\non\\
&&=\left[-a_{13}(z,-z) {B}(z) \bar{B}(z)+a_{13}(z,-z) D(z) \bar{D}(z)\right]|0\rangle\,.
\ee
Using the definition of $\tilde{\mathcal{A}}_{ij}(z)$, we obtain
\be
\mathcal{D}(z)|0\rangle&=&\frac{a_{10}(z,-z)}{a_3(z,-z)}\left[\tilde{\mathcal{A}}_{11}(z)+\tilde{\mathcal{A}}_{22}(z)\right]|0\rangle
+\frac{a_{13}(z,-z)}{a_3(z,-z)}\mathcal{B}(z)|0\rangle\non\\
&&+\frac{a_{14}(z,-z)}{a_3(z,-z)}D(z)\bar{D}(z)|0\rangle\,.
\ee
Defining
\be
\tilde{\mathcal{D}}(z)=\mathcal{D}(z)-\frac{a_{10}(z,-z)}{a_3(z,-z)}\left[\tilde{\mathcal{A}}_{11}(z)+\tilde{\mathcal{A}}_{22}(z)\right]
-\frac{a_{13}(z,-z)}{a_3(z,-z)}\mathcal{B}(z)\,,
\ee
we obtain
\be
\tilde{\mathcal{D}}(z)|0\rangle=\frac{a_{14}(z,-z)}{a_3(z,-z)}D(z)\bar{D}(z)|0\rangle\,.
\ee

The transfer matrix (\ref{transfer}) can be expressed in terms of
elements of the double-row monodromy matrix (\ref{doublerow})
\be
t(z)&=&\str_{a}\left\{R^{L}_{a}(z)\, U_{a}(z)\right\}\\
&=&K^+_1(z)\mathcal{A}_{11}(z)+K^+_2(z)\mathcal{A}_{21}(z)+K^+_3(z)\mathcal{A}_{12}(z)+K^+_4(z)\mathcal{A}_{22}(z)-\mathcal{D}(z)-\mathcal{B}(z)\non\\
&=&\bar{K}^+_1(z)\tilde{\mathcal{A}}_{11}(z)+\bar{K}^+_2(z)\tilde{\mathcal{A}}_{21}(z)+\bar{K}^+_3(z)\tilde{\mathcal{A}}_{12}(z)
+\bar{K}^+_4(z)\tilde{\mathcal{A}}_{22}(z)+\bar{K}^+_5(z)\tilde{\mathcal{D}}(z)+\bar{K}^+_6(z)\mathcal{B}(z)\,,\non
\ee
where
\be
&&\bar{K}^+_2(z)=K^+_2(z),\quad \bar{K}^+_3(z)=K^+_3(z),\quad
\bar{K}^+_5(z)=-1\,,\non\\
&&\bar{K}^+_1(z)=K^+_1(z)-\frac{a_{10}(z,-z)}{a_3(z,-z)}\,,\quad
\bar{K}^+_4(z)=K^+_4(z)-\frac{a_{10}(z,-z)}{a_3(z,-z)}\,,\non\\
&&\bar{K}^+_6(z)=-1-\frac{a_9(z,-z)}{a_3(z,-z)}\left[{K}^+_1(z)+{K}^+_4(z)\right]-\frac{a_{13}(z,-z)}{a_3(z,-z)}=-a_{12}(z,-z)\,.
\ee

It is now straightforward to verify from the above
results that the reference state is an eigenstate of the transfer
matrix, with eigenvalue
\be
\Lambda_{0}(z) &=&
\bar{K}^+_6(z)\prod_{k=1}^N a_3(z,z_k)a_3(z_k,-z)
+\bar{K}^+_5(z)a_{14}(z,-z) \prod_{k=1}^N a_{14}(z,z_k)a_{14}(z_k,-z) 
\non \\
&+& 2 \cos(2\theta) \cos^{2}\left[\frac{p(z)}{2}\right] \prod_{k=1}^N 
a_5(z,z_k)a_6(z_k,-z) \,.
\ee

\subsubsection{The action of the transfer matrix on the first-level
eigenstates}

Using the right reflection equation for the double-row monodromy
matrix (c.f. (\ref{RBYBE}))
\be
S_{12}(z_1,z_2)\, U_1(z_1)\, S_{21}(z_2,-z_1)\, U_2(z_2) =
U_2(z_2)\, S_{12}(z_1,-z_2)\, U_1(z_1)\, S_{21}(-z_2,-z_1)\,,
\ee
and the definitions of $\tilde{\mathcal{A}}_{ij}(z)$ and
$\tilde{\mathcal{D}}(z)$, we obtain -- after lengthy computations -- the following exchange relations:
\be
&&\mathcal{B}(z_1)\mathcal{B}_k(z_2)=\frac{a_3(z_2,z_1)a_6(z_1,-z_2)}{a_3(z_2,-z_1)a_6(-z_1,-z_2)}\mathcal{B}_k(z_2)\mathcal{B}(z_1)+{\rm{u.t.}},\label{relationBB}\\
&&\tilde{\mathcal{A}}_{a_1d_1}(z_1)\mathcal{B}_{c_1}(z_2)=\frac{r(z_1,-z_2)^{a_2b_1}_{a_1c_2}\bar{r}(-z_2,-z_1)^{d_1c_1}_{d_2b_1}}{a_5(z_1,z_2)a_6(z_2,-z_1)}
\mathcal{B}_{c_2}(z_2)\tilde{\mathcal{A}}_{a_2d_2}(z_1)+{\rm{u.t.}},\label{relationAB}\\
&&\tilde{\mathcal{D}}(z_1)\mathcal{B}_k(z_2)=\frac{a_{12}(z_1,-z_2)a_5(-z_2,-z_1)}{a_{14}(z_1,z_2)a_{5}(z_1,-z_2)}
\mathcal{B}_k(z_2)\tilde{\mathcal{D}}(z_1)+{\rm{u.t.}},\label{relationDB}\\
&&\vec{\mathcal{B}}(z_1)\otimes\vec{\mathcal{B}}(z_2)=-\frac{a_8(-z_2,-z_1)}{a_{14}(-z_2,-z_1)a_6(z_2,-z_1)}\left[a_{14}(z_1,-z_2)\mathcal{F}(z_2)\mathcal{B}(z_1)-
a_{14}(z_2,-z_1)\mathcal{F}(z_1)\mathcal{B}(z_2)\right]\vec{\xi}\non\\
&&\quad\quad-\frac{a_6(z_1,-z_2)}{a_3(z_1,z_2)a_6(z_2,-z_1)}\left\{\vec{\mathcal{B}}(z_2)\otimes\vec{\mathcal{B}}(z_1)+\frac{a_8(z_1,-z_2)}{a_6(z_1,-z_2)}
\mathcal{F}(z_2)\, \vec{\xi}\cdot\left[I\otimes\mathcal{A}(z_1)\right]\right\}\cdot\tilde{r}(-z_2,-z_1)\non\\
&&\quad\quad-\frac{a_8(z_2,-z_1)}{a_6(z_2,-z_1)}\mathcal{F}(z_1)\,
\vec{\xi}\cdot\left[I\otimes\mathcal{A}(z_2)\right]\,,
\label{relationBB2}
\ee
where
\be
&&\vec{\xi}=\left(
              \begin{array}{cccc}
                0, & 1, & -1, & 0 \\
              \end{array}
            \right)\,,\\
&&\mathcal{A}(z)=\left(
                   \begin{array}{cc}
                     \mathcal{A}_{11}(z) & \mathcal{A}_{12}(z) \\
                     \mathcal{A}_{21}(z) & \mathcal{A}_{22}(z) \\
                   \end{array}
                 \right)\,, \qquad \vec{\mathcal{B}}(z) =
		 \left(\mathcal{B}_{1}(z)\,, \mathcal{B}_{2}(z)\right) \,, \\
&&r(z_1,z_2)=\left(
          \begin{array}{cccc}
            h_1(z_1,z_2) &  &  &  \\
             & h_2(z_1,z_2) & h_3(z_1,z_2) &  \\
             & h_3(z_1,z_2) & h_2(z_1,z_2) &  \\
             &  &  & h_1(z_1,z_2) \\
          \end{array}
        \right)\,,\\
&&\bar{r}(z_1,z_2)=\left(
          \begin{array}{cccc}
            h_4(z_1,z_2) &  &  &  \\
             & h_5(z_1,z_2) & h_6(z_1,z_2) &  \\
             & h_6(z_1,z_2) & h_5(z_1,z_2) &  \\
             &  &  & h_4(z_1,z_2) \\
          \end{array}
        \right)\,,\\
&&\tilde{r}(z_1,z_2)=\left(
          \begin{array}{cccc}
            h_4(z_1,z_2) &  &  &  \\
             & h_6(z_1,z_2) & h_5(z_1,z_2) &  \\
             & h_5(z_1,z_2) & h_6(z_1,z_2) &  \\
             &  &  & h_4(z_1,z_2)
             \end{array}
        \right)\,,
\ee
with
\be
&&h_1(z_1,z_2)=a_1(z_1,z_2)+\frac{a_9(z_1,z_2)a_{10}(z_1,z_2)}{a_3(z_1,z_2)},\quad h_2(z_1,z_2)=a_{12}(z_1,z_2)\,,\non\\
&&h_3(z_1,z_2)=h_1(z_1,z_2)-h_2(z_1,z_2)\,,\quad
h_4(z_1,z_2)=a_1(z_1,z_2)\,,\\
&&h_5(z_1,z_2)=a_{12}(z_1,z_2)-\frac{a_7(z_1,z_2)a_8(z_1,z_2)}{a_{14}(z_1,z_2)},\quad h_6(z_1,z_2)=h_4(z_1,z_2)-h_5(z_1,z_2)\,, \non
\ee
and ``u.t.'' denotes so-called unwanted terms, which we do not explicitly write.

In terms of $u(z)$ (\ref{udef}), we can now write
\be
r(z_1, -z_2) &=& -i h_3(z_1, -z_2)\, R^{(2)}(u_{1}+u_{2}-i)\,, \non \\
\bar{r}(-z_2,-z_1) &=& -ih_6(-z_2,-z_1)\, R^{(2)}(u_{1}-u_{2})\,,
\label{rR}
\ee
where $u_{j} \equiv u(z_{j})$, and $R^{(2)}(u)$ is the familiar
spin-1/2 XXX $R$-matrix
\be
R^{(2)}(u) = u \id + i \Pi \,,
\ee
where $\id$ and $\Pi$ are the  $4 \times 4$ identity and permutation
matrices, respectively.

The first-level eigenvectors of the transfer matrix
have the general structure \cite{Martins:1997, Guan:1999}
\be
|\Phi_M(\lambda_1,\cdots,\lambda_M)\rangle=\vec\Phi_M(\lambda_1,\cdots,\lambda_M) \cdot \vec{F} |0\rangle,
\ee
where $\{\lambda_j\}$ are Bethe roots,
$\vec\Phi_M(\lambda_1,\cdots,\lambda_M)$ are $2^{M}$-dimensional
row-vectors whose components are operators, and $\vec{F}$ are
$c$-number coefficients.
The $\vec\Phi_M(\lambda_1,\cdots,\lambda_M)$ can be shown to satisfy
a recursion relation of the form\footnote{We note a typo in the third
term of Eq. (3.40) in \cite{Guan:1999}, and we thank X.-W. Guan for
correspondence on this point.}
\be
&&\vec\Phi_M(\lambda_1,\cdots,\lambda_M)=\vec{\mathcal{B}}(\lambda_1)\otimes\vec\Phi_{M-1}(\lambda_2,\cdots,\lambda_M)\\
&&\quad\quad+\sum_{j=2}^M \left[\vec{\xi}\otimes\mathcal{F}(\lambda_1)
\vec\Phi_{M-2}(\lambda_2,\cdots,\lambda_{j-1},\lambda_{j+1},\cdots,\lambda_M)\,
\mathcal{B}(\lambda_j)\right] g^{(M)}_{j-1}(\lambda_1,\cdots,\lambda_M)\non\\
&&\quad\quad-\sum_{j=2}^M \mathcal{F}(\lambda_1) \vec\Phi_{M-2}(\lambda_2,\cdots,\lambda_{j-1},\lambda_{j+1},\cdots,\lambda_M)
\otimes \left[\vec{\xi}\cdot (I\otimes\tilde{\mathcal{A}}(\lambda_j))\right]\, h^{(M)}_{j-1}(\lambda_1,\cdots,\lambda_M)\,, \non
\ee
for certain functions $g^{(M)}_{j-1}$ and $h^{(M)}_{j-1}$ whose
explicit expressions will not be needed here, and $\vec\Phi_0=1$.
In particular, $\vec\Phi_1(\lambda)=\vec{\mathcal{B}}(\lambda)$.

Let us define
\be
y_j=x^-(\lambda_j)\,, \qquad \tilde{u}_j =
\tfrac{g}{2}(y_j+\tfrac{1}{y_j})+\tfrac{i}{2}
\,, \qquad j=1,\cdots, M\,.
\label{yjdef}
\ee
By using the exchange relations (\ref{relationBB})-(\ref{relationDB})
and the values of $\tilde{\mathcal{A}}_{ij}(z)$,
$\tilde{\mathcal{D}}(z)$ and $\mathcal{B}(z)$
when acting on the reference state, we obtain
\be
&&t(z)|\Phi_M(\lambda_1,\cdots,\lambda_M)\rangle=\left\{\bar{K}^+_6(z)\prod_{j=1}^M\frac{a_3(\lambda_j,z)a_6(z,-\lambda_j)}{a_3(\lambda_j,-z)a_6(-z,-\lambda_j)}
\prod_{k=1}^Na_3(z,z_k)a_3(z_k,-z)\right.\non\\
&&\quad\quad\left.+\bar{K}^+_5(z)a_{14}(z,-z)\prod_{j=1}^M\frac{a_{12}(z,-\lambda_j)a_5(-\lambda_j,-z)}{a_{14}(z,\lambda_j)a_{5}(z,-\lambda_j)}
\prod_{k=1}^Na_{14}(z,z_k)a_{14}(z_k,-z)\right.\non\\
&&\quad\quad\left.+\cos^{2}\left[\frac{p(z)}{2}\right]
\prod_{j=1}^M-\frac{h_3(z,-\lambda_j)h_6(-\lambda_j,-z)}{a_5(z,\lambda_j)a_6(\lambda_j,-z)}\prod_{k=1}^Na_5(z,z_k)a_6(z_k,-z)\, t^{(2)}(z)\right\}
\non\\
&&\quad\quad\times|\Phi_M(\lambda_1,\cdots,\lambda_M)\rangle
+{\rm{u.t.}}\,, \label{nestedoffshell}
\ee
where $t^{(2)}(z)$ is the second-level nested transfer matrix with
inhomogeneities $\{\tilde{u}_j\}$ 
\be
t^{(2)}(z)=\tr_{0}\{K_{0}^{(2)+}(u)\,
T_{0}^{(2)}(u,\{\tilde{u}_j\})\, K_{0}^{(2)-}(u)\,
\widehat{T}_{0}^{(2)}(u,\{\tilde{u}_j\})\}\,,
\label{nestedtransfer}
\ee
with
\be
T_{0}^{(2)}(u,\{\tilde{u}_j\})&=&R^{(2)}_{0,1}(u+\tilde{u}_1-i)\cdots
R^{(2)}_{0,M}(u+\tilde{u}_M-i)\,,\\
\widehat{T}_{0}^{(2)}(u,\{\tilde{u}_j\})&=&R^{(2)}_{M,0}(u-\tilde{u}_M)\cdots R^{(2)}_{1,0}(u-\tilde{u}_1)\,, \\
K^{(2)-}(u)&=&\left(
                      \begin{array}{cc}
                        1 & 0 \\
                        0 & -1 \\
                      \end{array}
                    \right)\,,\\
K^{(2)+}(u)&=&\left(
                        \begin{array}{cc}
                          \cos(2\theta) & \sin(2\theta) \\
                          \sin(2\theta) & -\cos(2\theta) \\
                        \end{array}
                      \right) \,. \label{Kplus}
\ee
We remind the reader that $u=u(z)$ is given by  (\ref{udef}), and therefore
\be
u(z) =
\tfrac{g}{4}\left[x^{+}(z)+\tfrac{1}{x^{+}(z)}+x^{-}(z)+\tfrac{1}{x^{-}(z)}\right] \,.
\ee

\subsection{Off-diagonal Bethe ansatz}\label{sec:ODBA}

In view of (\ref{nestedoffshell}), in order to determine the
eigenvalues of the transfer matrix (\ref{transfer}), it now remains to
diagonalize the nested transfer matrix $t^{(2)}(z)$. We recognize
the latter as the transfer matrix of an open spin-1/2 XXX chain of
length $M$ with non-diagonal boundary terms. Therefore, using  the
off-diagonal Bethe ansatz \cite{Cao:2013qxa, Wang2015} 
(see also \cite{Nepomechie:2013ila}), we can immediately write down an
expression for the corresponding eigenvalues.

Indeed, let us introduce the following functions
\be
a^{(2)}(u)&=&\frac{2u-i}{2u}\prod_{j=1}^M (u-\tilde{u}_j+i)(u+\tilde{u}_j)\,, \\
d^{(2)}(u)&=&\frac{2u+i}{2u}\prod_{j=1}^M
(u-\tilde{u}_j)(u+\tilde{u}_j-i)\,.
\ee
According to the off-diagonal Bethe ansatz, the eigenvalues of $t^{(2)}(z)$
can be given by 
\be
\Lambda^{(2)}(z)&=&a^{(2)}(u)\frac{Q_{2}(u-i)}{Q_{2}(u)}
+d^{(2)}(u)\frac{Q_{2}(u+i)}{Q_{2}(u)} \label{Lambdanested}\\
&+& \frac{2\left[\cos(2\theta)-1\right]}{Q_{2}(u)}
\prod_{j=1}^M (u-\tilde{u}_j)(u+\tilde{u}_j-i)(u-\tilde{u}_j+i)(u+\tilde{u}_j)\,,\non
\ee
where the polynomial $Q_{2}(u)$ is parameterized by $M$ Bethe roots
$\{w_j\}$
\be
Q_{2}(u)=\prod_{j=1}^M (u-w_j)(u+w_j) \,.
\ee
One can recognize (after multiplying both sides by $Q_{2}(u)$)
that (\ref{Lambdanested}) is a T-Q equation with an additional inhomogeneous
term.

\subsection{Eigenvalues and Bethe equations}

Combining the results (\ref{nestedoffshell}) and
(\ref{Lambdanested}), we conclude that the eigenvalues of the
transfer matrix $t(z)$ (\ref{transfer}) are given by 
\be
&&\Lambda(z)=-a_{12}(z,-z)e^{-iMp(z)}\prod_{j=1}^M
\frac{[x^{+}(z)-y_j][x^{+}(z)+y_j]}{[x^{-}(z)-y_j][x^{-}(z)+y_j]}
\prod_{k=1}^Na_3(z,z_k)a_3(z_k,-z)\non\\
&&\quad\quad-a_{14}(z,-z)e^{iMp(z)}
\prod_{j=1}^M\frac{[x^-(z)y_j-1][x^-(z)y_j+1]}{[x^+(z)y_j-1][x^+(z)y_j+1]}
\prod_{k=1}^Na_{14}(z,z_k)a_{14}(z_k,-z)\non\\
&&\quad\quad+\cos^{2}\left[\frac{p(z)}{2}\right]
\prod_{k=1}^Na_5(z,z_k)a_6(z_k,-z)
\Bigg\{ \label{Lambaresult}\\
&&\quad\quad \frac{2u-i}{2u}
e^{-iMp(z)}\frac{Q_{2}(u-i)}{Q_{2}(u)}
\prod_{j=1}^M \frac{[x^{+}(z)-y_j][x^{+}(z)+y_j]}{[x^{-}(z)-y_j][x^{-}(z)+y_j]}\non\\
&&\quad\quad +  \frac{2u+i}{2u}
e^{iMp(z)}\frac{Q_{2}(u+i)}{Q_{2}(u)}
\prod_{j=1}^M\frac{[x^-(z)y_j-1][x^-(z)y_j+1]}{[x^+(z)y_j-1][x^+(z)y_j+1]}\non\\
&&\quad\quad+\frac{2[\cos(2\theta)-1]}{Q_{2}(u)}
\prod_{j=1}^M
\big(\frac{g^{2}}{4}\big)\frac{[x^+(z)+y_j][x^+(z)-y_j][x^-(z)y_j+1][x^-(z)y_j-1]}{x^+(z)x^-(z)y^2_j}\Bigg\}\,.\non
\ee
The requirement that $\Lambda(z)$ should not have any poles leads to
the following Bethe equations\footnote{It should also
be possible to obtain the Bethe equations from the cancellation of the
unwanted terms that appear when the transfer matrix acts on an
off-shell Bethe state; however, we have not determined the complete
off-shell equation. Such an off-shell equation has been found
recently for the XXZ chain \cite{Avan:2015}.}
\be
&&\prod_{l=1}^N\frac{x^-(z_l)+y_j}{x^+(z_l)+y_j}\frac{x^+(z_l)-y_j}{x^-(z_l)-y_j}
\frac{Q_{2}(\tilde{u}_j)}{Q_{2}(\tilde{u}_j-i)}=1\,, \qquad
j=1,\cdots,M\,, \label{BAE1} \\
&&a^{(2)}(w_k)Q_{2}(w_k-i)+d^{(2)}(w_k)Q_{2}(w_k+i) \label{BAE2}  \\
&&+2\left[\cos(2\theta)-1\right]
\prod_{j=1}^M
(w_k-\tilde{u}_j)(w_k+\tilde{u}_j-i)(w_k-\tilde{u}_j+i)(w_k+\tilde{u}_j) =0\,,
\quad k=1,\cdots,M \,. \non
\ee

These results can be reexpressed more succinctly
by using the shorthand notation of \cite{Gromov:2009tv, Bajnok:2012xc}
\be
R^{(\pm)}(z)&=&\prod_{i=1}^{N}\left(x(z)-x^{\mp}(z_{i})\right)\left(x(z)+x^{\pm}(z_{i})\right)\,, \non \\
B^{(\pm)}(z)&=& R^{(\pm)}(z+\omega_{2}) =
\prod_{i=1}^{N}\left(\frac{1}{x(z)}-x^{\mp}(z_{i})\right)\left(\frac{1}{x(z)}+x^{\pm}(z_{i})\right)\,.
\label{RBdefs}
\ee
For example, the expression $R^{(-)-}(z)$ should be understood to mean
\be
R^{(-)-}(z) &=&
\prod_{i=1}^{N}\left(x^{-}(z)-x^{+}(z_{i})\right)\left(x^{-}(z)+x^{-}(z_{i})\right) \,. \non
\ee
Similarly,
\be
B_{1}R_{3}(z)&=&\prod_{j=1}^{M}\left(x(z)-y_{j}\right)\left(x(z)+y_{j}\right) \,, \non \\
R_{1}B_{3}(z)&=& B_{1}R_{3}(z+\omega_{2}) =
\prod_{j=1}^{M}\left(\frac{1}{x(z)}-y_{j}\right)\left(\frac{1}{x(z)}+y_{j}\right)  \,.
\ee
Moreover, if $f(u)$ is any function of $u$, then $f^{\pm}=f(u\pm
\tfrac{i}{2})\,, f^{\pm\pm}= f(u\pm i)$.

In terms of this notation, the eigenvalues (\ref{Lambaresult}) are
given by 
\be
&&\Lambda(z)=e^{i(N-M+1)p}\frac{1}{R^{(+)+} B^{(-)+}}\Bigg\{
-\rho_{1} R^{(-)-} B^{(-)+} \frac{(B_{1} R_{3})^{+}}{(B_{1} R_{3})^{-}}
-\rho_{2} R^{(+)-} B^{(+)+} \frac{(R_{1} B_{3})^{-}}{(R_{1} B_{3})^{+}} \non\\
&&\quad\quad+ \tfrac{1}{2}(\rho_{1}+\rho_{2}) R^{(+)-} B^{(-)+} \Bigg[
 \frac{u^{-}}{u} \frac{(B_{1} R_{3})^{+}}{(B_{1} R_{3})^{-}}
\frac{Q_{2}^{--}}{Q_{2}}+
 \frac{u^{+}}{u} \frac{(R_{1} B_{3})^{-}}{(R_{1} B_{3})^{+}}
\frac{Q_{2}^{++}}{Q_{2}}\non\\
&&\quad\quad+2[\cos(2\theta)-1]\frac{(R_{1} B_{3})^{-} (B_{1}
R_{3})^{+}}{Q_{2}} \prod_{j=1}^{M} \left(-\tfrac{g^{2}}{4
y_{j}^{2}}\right) \Bigg]\Bigg\}\,,
\label{Lambda2}
\ee
where
\be
\rho_{1}(z)=\frac{(1+(x^{-})^{2})(x^{-}+x^{+})}{2 x^{+} (1+ x^{-} x^{+})} \,, \quad
\rho_{2}(z) = \rho_{1}(-z-\omega_{2}) = \frac{x^{-}
(1+(x^{+})^{2})(x^{-}+x^{+})}{2 (x^{+})^{2} (1+ x^{-} x^{+})} \,.
\label{ab}
\ee
The corresponding Bethe equations are 
\be
&&\frac{R^{(-)-}}{R^{(+)-}}\frac{Q_{2}}{Q_{2}^{--}}\Bigg\vert_{z=\lambda_{j}} = 1\,, \qquad j=1, \ldots, M\,, \label{BAE1a} \\
&&\left[\tfrac{u^{-}}{u}Q_{1}^{+} Q_{2}^{--} + \tfrac{u^{+}}{u}Q_{1}^{-}
Q_{2}^{++}
+2[\cos(2\theta)-1]\, Q_{1}^{+} Q_{1}^{-}\right]\Bigg\vert_{u=w_{k}}  =0\,,
\quad k=1,\cdots,M \,,  \label{BAE2a}
\ee
where \footnote{We recall the definitions (\ref{yjdef}) and also note
the identity
\be
(B_{1} R_{3})^{\pm} (R_{1} B_{3})^{\pm} = Q_{1}^{\pm} \prod_{j=1}^{M}
\left(-\tfrac{4 y_{j}^{2}}{g^{2}}\right) \,. \non
\ee}
\be
Q_{1}(u) = \prod_{j=1}^M
(u+\tfrac{i}{2}-\tilde{u}_j)(u-\tfrac{i}{2}+\tilde{u}_j)
\,.
\ee
The Bethe equations (\ref{BAE1a}) and (\ref{BAE2a}) are
equivalent to (\ref{BAE1}) and (\ref{BAE2}), respectively.

For $\theta=0$, the last (``inhomogeneous'') term in
(\ref{Lambda2}) vanishes; and we see (using
$\frac{\rho_{2}}{\rho_{1}}=\frac{u^{+}}{u^{-}}$) that
our result  (\ref{Lambda2}) for the transfer matrix
eigenvalue is consistent with the $sl(2)$ grading result
(C.8) in \cite{Bajnok:2012xc}.

Interestingly, for $\theta=\pi/2$, the inhomogeneous term does {\em
not} vanish, even though the boundary $S$-matrices are diagonal for
this case (see (\ref{boundSmat}), (\ref{boundSmatelem}) and (\ref{Kplus}))!
This is the price we pay for having an expression for $\Lambda(z)$ that
is analytic in $\theta$. 
An alternative expression is \footnote{This expression corresponds to 
choosing a different parametrization for the T-Q equation 
(\ref{Lambdanested}), with $a^{(2)}(u)$ and $d^{(2)}(u)$ rescaled by 
$s(\theta)$ and with a corresponding modification of the 
inhomogeneous term. One can show that both parametrizations satisfy the necessary requirements of crossing symmetry, 
initial condition, asymptotic behavior,  and functional relation \cite{Cao:2013qxa, Wang2015}.}
\be
&&\Lambda(z)=e^{i(N-M+1)p}\frac{1}{R^{(+)+} B^{(-)+}}\Bigg\{
-\rho_{1} R^{(-)-} B^{(-)+} \frac{(B_{1} R_{3})^{+}}{(B_{1} R_{3})^{-}}
-\rho_{2} R^{(+)-} B^{(+)+} \frac{(R_{1} B_{3})^{-}}{(R_{1} B_{3})^{+}} \non\\
&&\quad\quad+ \tfrac{1}{2}(\rho_{1}+\rho_{2}) R^{(+)-} B^{(-)+} \Bigg[
s(\theta) \frac{u^{-}}{u} \frac{(B_{1} R_{3})^{+}}{(B_{1} R_{3})^{-}}
\frac{Q_{2}^{--}}{Q_{2}}+
s(\theta) \frac{u^{+}}{u} \frac{(R_{1} B_{3})^{-}}{(R_{1} B_{3})^{+}}
\frac{Q_{2}^{++}}{Q_{2}}\non\\
&&\quad\quad+2[\cos(2\theta)-s(\theta)]\frac{(R_{1} B_{3})^{-} (B_{1}
R_{3})^{+}}{Q_{2}} \prod_{j=1}^{M} \left(-\tfrac{g^{2}}{4
y_{j}^{2}}\right) \Bigg]\Bigg\}\,, \non
\ee
where 
\be
s(\theta) = \frac{\cos(2\theta)}{|\cos(2\theta)|} =
\left\{ \begin{array}{rr}
1 & 0 \le \theta < \frac{\pi}{4} \\
-1 &  \frac{\pi}{4} < \theta \le \frac{\pi}{2}
\end{array} \right. \non
\,.
\ee
(For $\theta=\pi/4$, either $s=+1$ or $s=-1$ can be chosen.)
The inhomogeneous term in this expression does vanish for both
$\theta=0$ and $\theta=\pi/2$; and for $\theta=\pi/2$,
this result for the transfer matrix
eigenvalue is consistent with the duality transformation of the
result (3.20) in \cite{Bajnok:2013wsa}.

\subsection{Degeneracy and multiplicity}\label{sec:degmult}

The degeneracy of the transfer matrix eigenvalue (\ref{Lambda2}) corresponding to
a given solution of the Bethe equations (\ref{BAE1a})-(\ref{BAE2a}), as well as the number of
such solutions (multiplicity),  can be inferred
from the unbroken $su(2)$ symmetry (\ref{su2}) of the transfer
matrix. \footnote{We assume here that there are no pathologies, such
as singular solutions of the Bethe equations, or accidental spectrum
degeneracy.}

Indeed, we expect (see e.g. \cite{Martins:1997, Essler:2005})
that the Bethe states are $su(2)$ lowest-weight states, with
\be
s = -m = \tfrac{1}{2}(N-M) \,,
\ee
where $s(s+1)$ is the eigenvalue of $\vec{S}^{2}$, and $m$ is the
eigenvalue of $S^{z}$. Since $s \ge 0$, it follows that $M$ can take the values $0, 1,
\ldots, N$. Hence, for given values of $N$ and $M$, we expect that the degeneracy
${\cal D}(N,M)$ of the corresponding eigenvalue is given by
\be
{\cal D}(N,M) = 2s +1 = N-M+1 \,.
\label{degeneracy}
\ee

For one site, the decomposition of the 4-dimensional
vector space into $su(2)$  representations
is given by ${\bf 0} \oplus {\bf 0}
\oplus {\bf \tfrac{1}{2}}$. For $N$ sites, the decomposition
of the space of states into a direct sum of  $su(2)$ irreducible
representations can be easily determined using the Clebsch-Gordan theorem
\be
\left({\bf 0} \oplus {\bf 0} \oplus {\bf
\tfrac{1}{2}}\right)^{\otimes N} = \bigoplus_{s=0}^{N/2} n_{s}\, {\bf s} \,,
\label{CG}
\ee
where $n_{s}$ is the multiplicity of spin $s$.  With the help of the
multinomial theorem, an explicit expression for $n_{s}$ can be derived
\be
n_{s}=\sum_{\scriptstyle{k_{1}, k_{2}, k_{3}=0}\atop \scriptstyle{k_{1}+k_{2}+k_{3}=N}}^{N}\frac{N!}{k_{1}! k_{2}! k_{3}!}
d_{s}(k_{3})\,,
\label{ns}
\ee
where
\be
d_{s}(k_{3}) =  {k_{3}\choose \frac{k_{3}}{2}-s} - {k_{3}\choose
\frac{k_{3}}{2}-s-1} \,,
\ee
and ${n\choose m} = \frac{n!}{(n-m)! m!}$ is defined to be 0 if $m$
is outside of the interval $[0,n]$ or if $m$  is not an integer.
Hence, for given values of $N$ and $M$,
we expect that the number of solutions ${\cal N}(N,M)$ of the Bethe equations is given by
\be
{\cal N}(N,M) = n_{s}\Big\vert_{s=\tfrac{1}{2}(N-M)} \,,
\label{numbersltns}
\ee
where $n_{s}$ is given by (\ref{ns}). We have verified that the
expressions (\ref{degeneracy}) and  (\ref{numbersltns}) satisfy
the completeness constraint
\be
\sum_{M=0}^{N} {\cal D}(N,M)\, {\cal N}(N,M) = 4^{N}\,.
\ee

\subsection{Numerical checks}

We have numerically checked our Bethe ansatz solution
(\ref{Lambda2})-(\ref{BAE2a}), as well as the formulas for
degeneracies (\ref{degeneracy}) and multiplicities (\ref{numbersltns}),
for $N=1$ and $N=2$.

For $N=1$, we expect according to (\ref{numbersltns}) one solution
with $M=0$ (namely, the trivial solution with no Bethe roots,
corresponding to the reference state (\ref{reference})), and two
solutions with $M=1$. We indeed find these solutions, as shown in
Table \ref{table:N1}.
The corresponding eigenvalues obtained from (\ref{Lambda2}) match with the 4 eigenvalues obtained by direct
diagonalization of the transfer matrix (\ref{transfer}).


\begin{table}[htb]
   \small
 \centering
 \begin{tabular}{|l|c|c|c|}\hline
   $M$ & $\{ y_{j} \}$ & $\{ w_{k} \}$ & degeneracy \\
   \hline
    0 & - & - & 2 \\
    1 & 29.67201576134 & 4.98947370172 & 1 \\
    1 & 37.59406315269 & 4.98947370172 & 1 \\
   \hline
   \end{tabular}
  \caption{\small Solutions of the Bethe equations
  (\ref{BAE1a})-(\ref{BAE2a}) and degeneracies (\ref{degeneracy}) of the corresponding
  eigenvalues  (\ref{Lambda2}) for $N=1$ with $g=0.3$, $\theta = 0.7$,
  $z_{1}=0.1$.}
  \label{table:N1}
\end{table}

Similarly, for $N=2$,  we expect (\ref{numbersltns})  one solution
with $M=0$, four solutions with $M=1$, and five solutions
with $M=2$. We indeed find these solutions, as shown in Table
\ref{table:N2}.
The corresponding eigenvalues obtained from (\ref{Lambda2}) match
with the 16 eigenvalues obtained by direct
diagonalization of the transfer matrix.


\begin{table}[htb]
   \small
 \centering
 \begin{tabular}{|l|c|c|c|}\hline
   $M$ & $\{ y_{j} \}$ & $\{ w_{k} \}$ & degeneracy \\
   \hline
    0 & - & - & 3 \\
    1 & 1.688644387948 & 0.522719047641  & 2 \\
    1 & 5.425599080922 & 0.735491947795 $i$ & 2 \\
    1 & 8.578126668210 &  1.731865667315 & 2 \\
    1 & 21.148232916045 & 2.466473986963 & 2\\
    2 &  5.114946745748, 18.101713927816 & 0.461647669632, 1.336004479377 & 1 \\
    2 &  2.201092869446, 16.716623804005 & 0.617999330346, 1.271260542723 & 1 \\
    2 &   1.035542549912,  7.917752654460 & 1.033924690882 $\pm$
    0.264550555698 $i$  & 1 \\
    2 &  10.157304730304 $\pm$ 2.8436106714245 $i$ & 0.880379764515, 1.105874319271 & 1 \\
    2 &  0.129605338411,  6.858626722293 & 1.783773124400,
    1.088034934890 $i$ & 1 \\
   \hline
   \end{tabular}
  \caption{\small Solutions of the Bethe equations
  (\ref{BAE1a})-(\ref{BAE2a}) and degeneracies (\ref{degeneracy}) of the corresponding
  eigenvalues  (\ref{Lambda2}) for $N=2$ with $g=0.3$, $\theta = 0.7$,
  $z_{1}=0.8$, $z_{2}=0.4$.}
  \label{table:N2}
\end{table}

In short, we have verified that our Bethe ansatz solution correctly
gives the complete set of eigenvalues of the transfer matrix for $N=1$ and $N=2$.



\section{Discussion}\label{sec:discussion}

For the transfer matrix (\ref{transfer}) of the $Y_{\theta}-Y$ system,
we have determined the exact eigenvalues (\ref{Lambda2}) in terms of
solutions of a corresponding set of Bethe equations
(\ref{BAE1a})-(\ref{BAE2a}).  We have checked this result numerically
for small system size.

The $Y_{\theta}=0$ boundary $S$-matrix (\ref{LBSM})
is one of the few known integrable AdS/CFT boundary $S$-matrices with a free
parameter.  (Other examples are discussed in \cite{Murgan:2008fs, 
Prinsloo:2015apa}.)
The present work represents the first time in the AdS/CFT context
that an open-chain transfer matrix with a non-diagonal boundary
$S$-matrix is diagonalized.

We hope to use this result in a future publication to compute
asymptotic energies and finite-size
corrections for one-particle states, as a function of the
angle $\theta$.  Such corrections have already been computed for the
special (diagonal) cases $Y-Y$ $(\theta=0)$ and $\bar{Y}-Y$
$(\theta=\pi/2)$ in \cite{Bajnok:2010ui, Bajnok:2012xc} and
\cite{Bajnok:2013wsa}, respectively.  The latter system is noteworthy
for the presence of tachyons in its spectrum.

We expect that similar techniques can also be used to analyze other
integrable cases with non-diagonal boundary $S$-matrices.

\section*{Acknowledgments}
RN thanks Zoltan Bajnok and Laszlo Palla for valuable discussions
and comments on a preliminary draft, Xi-Wen Guan for correspondence, and
the Institute of Physics - Chinese Academy of Sciences for its
kind hospitality.
Financial support from the NSFC under Grant Nos.  11174335, 11375141,
11374334, 11434013 and 11425522, the National Program for Basic
Research of MOST, BCMIIS, and the Strategic Priority Research Program
of the CAS is gratefully acknowledged.
The work of RN was supported in part by the National Science
Foundation under Grant PHY-1212337, and by a Cooper fellowship.

\appendix

\section{Generating functional for higher transfer matrices}

In the body of this paper, we have focused on a transfer matrix $t(z)$
(\ref{transfer}) whose auxiliary space belongs to the fundamental (4-dimensional)
representation of $su(2|2)$. This transfer matrix is only
the first member of an infinite hierarchy of commuting transfer matrices
$T_{a,s}$ (with $T_{1,1}=t(z)$) whose auxiliary spaces belong to
rectangular representations of $su(2|2)$, and which satisfy the
Hirota equation \cite{Gromov:2009tv}
\be
T^{+}_{a,s}\, T^{-}_{a,s} = T_{a+1,s}\, T_{a-1,s} + T_{a,s+1}\,
T_{a,s-1} \,.
\label{Hirota}
\ee
We propose here a generating functional for the eigenvalues of these
transfer matrices (which we also denote by $T_{a,s}$), which are
useful for computing finite-size corrections (see e.g.
\cite{Gromov:2009tv, Bajnok:2012xc, Bajnok:2013wsa}). This generating functional
is a generalization of the one proposed in \cite{Nepomechie:2013ila}
for the XXX chain with nondiagonal boundary terms.

In order to streamline the notation, we rewrite the eigenvalue result (\ref{Lambda2})
as
\be
T_{1,1} = h\, \hat{T}_{1,1} \,, \qquad \hat{T}_{1,1} = -A - B + G + H + C \,,
\label{T11}
\ee
where $h$ is a normalization factor
\be
h = \rho_{1} \left(\frac{x^{+}}{x^{-}}\right)^{N-M+1}
\frac{R^{(+)-}}{R^{(+)+}} \,,
\ee
and
\be
A &=& \frac{R^{(-)-}}{R^{(+)-}}  \frac{(B_{1} R_{3})^{+}}{(B_{1}
R_{3})^{-}} \,, \non \\
B &=& \frac{u^{+}}{u^{-}} \frac{B^{(+)+}}{B^{(-)+}}
 \frac{(R_{1} B_{3})^{-}}{(R_{1} B_{3})^{+}} \,, \non \\
G &=& \frac{(B_{1} R_{3})^{+}}{(B_{1} R_{3})^{-}}
\frac{Q_{2}^{--}}{Q_{2}} \,, \non \\
H &=& \frac{u^{+}}{u^{-}} \frac{(R_{1} B_{3})^{-}}{(R_{1} B_{3})^{+}}
\frac{Q_{2}^{++}}{Q_{2}}\,, \non\\
C &=& [\cos(2\theta)-1]\left(1 +  \frac{u^{+}}{u^{-}} \right)
\frac{(R_{1} B_{3})^{-} (B_{1} R_{3})^{+}}{Q_{2}} \prod_{j=1}^{M} \left(-\tfrac{g^{2}}{4
y_{j}^{2}}\right)   \,.
\label{ABGHC}
\ee
We propose that the generating functional for antisymmetric
representations is given by
\be
W^{-1} &=& (1-{\cal D} A {\cal D})^{-1} \left[1-{\cal D}(G + H + C) {\cal D}
+ {\cal D} G {\cal D}^{2} H {\cal D} \right] (1-{\cal D} B {\cal
D})^{-1} \non \\
&=& \sum_{a=0}^{\infty} (-1)^{a} {\cal D}^{a}\,  \hat{T}_{a,1}\, {\cal
D}^{a} \,,
\label{genfunc}
\ee
where ${\cal D} = e^{-\frac{i}{2}\partial_{u}}$ implying ${\cal D} f
= f^{-} {\cal D}$, with
\be
T_{a,1} = h^{[a-1]} h^{[a-3]} \cdots h^{[3-a]} h^{[1-a]}\,
\hat{T}_{a,1} \,,
\ee
where $f^{[\pm n]}=f(u\pm \frac{in}{2})$.
By expanding both sides of (\ref{genfunc}), we obtain:
$\hat{T}_{0,1}=1$, the result in (\ref{T11}) for $\hat{T}_{1,1}$, and
\be
\hat{T}_{2,1} =  G^{+} H^{-}
- A^{+}\left(G^{-}+ H^{-}+ C^{-} - A^{-}\right) - \left(G^{+}+ H^{+}+
C^{+} -  A^{+} - B^{+}\right)  B^{-} \,,
\label{T21}
\ee
etc.

As a check on our proposal, we observe that for $\theta=0$ (and 
therefore $C=0$),
the factor in square brackets in (\ref{genfunc}) factorizes
\be
1-{\cal D}(G + H) {\cal D} + {\cal D} G {\cal D}^{2} H {\cal D} =
 (1-{\cal D} G {\cal D})  (1-{\cal D} H {\cal D}) \,,
\ee
and therefore the generating functional (\ref{genfunc}) reduces to
\be
W^{-1}\Big\vert_{\theta=0} = (1-{\cal D} A {\cal D})^{-1} (1-{\cal D} G {\cal D})  (1-{\cal D} H {\cal D}) (1-{\cal D} B {\cal
D})^{-1} \,,
\ee
which coincides with the result (C.10) in \cite{Bajnok:2012xc}.

A further check on our proposal is provided by the
special case $N=M=0$ and generic angle $\theta$. For this case, the
expressions in (\ref{ABGHC})
reduce to
\be
A = 1\,, \quad B = \frac{u^{+}}{u^{-}} \,, \quad G = 1\,,
\quad H =  \frac{u^{+}}{u^{-}} \,,
\quad C =  [\cos(2\theta)-1]\left(1 +  \frac{u^{+}}{u^{-}}
\right) \,,
\ee
and hence
\be
G + H + C = \cos(2\theta) \left(1 +  \frac{u^{+}}{u^{-}}\right) \,.
\ee
The generating functional (\ref{genfunc}) therefore reduces to
\be
W^{-1}\Big\vert_{N=M=0}= (1-{\cal D}^{2})^{-1} \left[1-\cos(2\theta){\cal D}(1 +  \tfrac{u^{+}}{u^{-}}) {\cal D}
+ {\cal D}^{3} \tfrac{u^{+}}{u^{-}} {\cal D} \right] (1-{\cal D} \tfrac{u^{+}}{u^{-}} {\cal
D})^{-1} \,,
\ee
which coincides with the result given by (E.13) and (E.17) in
\cite{Bajnok:2013wsa}. Moreover, for $N \ne 0$ but still $M=0$, we
obtain
\be
W^{-1}\Big\vert_{M=0} = (1-{\cal D}^{2} \tfrac{R^{(-)}}{R^{(+)}} )^{-1}
\left[1-\cos(2\theta){\cal D}(1 +  \tfrac{u^{+}}{u^{-}}) {\cal D}
+ {\cal D}^{3} \tfrac{u^{+}}{u^{-}} {\cal D} \right] (1-
\tfrac{B^{(+)}}{B^{(-)}}{\cal D} \tfrac{u^{+}}{u^{-}} {\cal
D})^{-1} \,,
\ee
which coincides with (E.21) in \cite{Bajnok:2013wsa}.

The generating functional for symmetric representations is given by
the inverse of (\ref{genfunc}),
\be
W &=& (1-{\cal D} B {\cal D}) \left[1-{\cal D}(G + H + C) {\cal D}
+ {\cal D} G {\cal D}^{2} H {\cal D} \right]^{-1} (1-{\cal D} A {\cal
D}) \non \\
&=& \sum_{s=0}^{\infty} {\cal D}^{s}\,  \hat{T}_{1,s}\, {\cal
D}^{s} \,,
\label{genfuncsymm}
\ee
with
\be
T_{1,s} = h^{[s-1]} h^{[s-3]} \cdots h^{[3-s]} h^{[1-s]}\,
\hat{T}_{1,s} \,.
\ee
By expanding both sides of (\ref{genfuncsymm}), we obtain:
$\hat{T}_{1,0}=1$, the result in (\ref{T11}) for $\hat{T}_{1,1}$, and
\be
\hat{T}_{1,2} =
\left(G^{+}+ H^{+}+ C^{+} - B^{+}\right)\left(G^{-}+ H^{-}+ C^{-} - A^{-}\right) - G^{+} H^{-} \,,
\label{T12}
\ee
etc. As a consistency check, it is now straightforward to verify the
Hirota equation (\ref{Hirota}) (which holds also for the renormalized
quantities $\hat{T}_{a,s}$) with $a=s=1$ using the results
(\ref{T11}), (\ref{T21}) and (\ref{T12}).


\providecommand{\href}[2]{#2}\begingroup\raggedright\endgroup

\end{document}